\documentclass{sigcomm-alternate}
\usepackage{amsfonts,latexsym,amsmath,amstext,amssymb,verbatim,epsfig,psfrag}
\usepackage{colordvi,pstcol}
\usepackage{graphicx,psboxit}
\usepackage{psfrag}

\def\ind{{\bf 1}}
\newtheorem{remark}{Remark}
\newtheorem{theorem}{Theorem}
\newtheorem{lemma}{Lemma}

\begin{document}
\title{Optimized on-line computation of PageRank algorithm}

\numberofauthors{1}
\author{
   \alignauthor Dohy Hong\vspace{2mm}\\
   \affaddr{Alcatel-Lucent Bell Labs}\\
   \affaddr{Route de Villejust}\\
   \affaddr{91620 Nozay, France}\\
   \email{\normalsize dohy.hong@alcatel-lucent.com}
}

\date{\today}
\maketitle

\begin{abstract}
In this paper we present new ideas to accelerate the computation of the eigenvector
of the transition matrix associated to the PageRank algorithm.
New ideas are based on the decomposition of the matrix-vector product that
can be seen as a fluid diffusion model, associated to new algebraic equations.
We show through experimentations on synthetic data and on real data-sets how much this
approach can improve the computation efficiency.
\end{abstract}
\begin{psfrags}
\section{Introduction}\label{sec:intro}
In this paper, we are interested by the computation issue of the solution
of the PageRank equation. 
PageRank is a link analysis algorithm that been initially introduced by \cite{page} and used
by the Google Internet search engine, that assigns a numerical value to each element of a hyper-linked 
set of nodes, such as the World Wide Web.
The algorithm may be applied to any collection of entities (nodes) that are linked through
directional relationships. 
The numerical value that it assigns to each node is called the PageRank of the node and
as we will see below the rank value of the node is associated to a eigenvector problem.

The complexity of the problem for computing the eigenvector of a matrix increases rapidly with 
the dimension of the vector space.
Efficient, accurate methods to compute eigenvalues and eigenvectors of arbitrary matrices
are in general a difficult problem (cf. power iteration \cite{mises}, QR algorithm \cite{francis, kub}). 

In the particular case of PageRank equation, several specific solutions were proposed and analyzed
\cite{deep, bian} including power method \cite{page} with adaptation \cite{kamvar} or
extrapolation \cite{haveliwala, kamvar2, brezinski},
iterative aggregation/disaggregation method \cite{lang2, kirkland, marek}, 
adaptive on-line method \cite{serge}, etc.

The approach proposed here is an improvement idea partially inspired from the algorithm
proposed in \cite{serge}. We also proposed an algebraic proof of Lemma 2.3 in \cite{serge}.
This approach can be also compared to the Gauss-Seidel iteration (cf. \cite{Saad}): 
the Gauss-Seidel iteration is known to be faster than the Jacobi iteration (cf. \cite{Arasu02pagerankcomputation}). 
However the approach can not be distributed because of the constraint in the order of the iteration
(\cite{Kohlschutter06efficientparallel}). 
In our approach,
the computation of each iteration uses the elements of the previous integrating the last 
update per component as with Gauss-Seidel method, as opposed to the power iteration (or to Jacobi method)
where the computation of the whole vector is based on the previous vector, without taking into
account the update (or partial update) at vector entry level.

As we will show below, our approach has the advantage of being iteration order independent and 
can be very naturally deployed using an asynchronous distributed computation.

\begin{figure}[htbp]
\centering
\includegraphics[width=\linewidth]{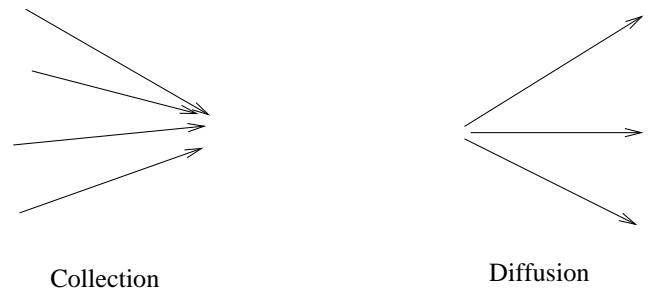}
\caption{Intuition: collection vs diffusion.}
\label{fig:cd}
\end{figure}

Finally, if one can associate the Gauss-Seidel method to an operation of {\em collection}
(one entry of vector is updated based on the previous vector based on the incoming links),
our approach consists in an operation of {\em diffusion} (the fluid diffusion from one entry
of the vector consists in updating all children nodes following the outgoing links) (cf. Figure \ref{fig:cd}).

The model is introduced in Section \ref{sec:model}.
Some theoretical results are presented in Section \ref{sec:theory}.
Finally we compared the computation cost of our approach to existing methods
through several data-sets in Section \ref{sec:simu}.

\section{Model}\label{sec:model}
\subsection{Notations}
We consider a non-negative matrix $P$ of size $N\times N$ such that each column sums up to one.
In particular, we can associate such a matrix to
a Markov chain on a state of size $N$, where $P$ would be
the transition matrix with $P_{ij} = p(i,j)$
the transition probability from state $j$ to state $i$.
In the following, we will also call $i$ as a node (from web graph
and PageRank context).

The fact that each column of $P$ sums up to one means that
in the Markov chain, each state $j$ has a positive probability
to jump to at least one state $i$
(if such a condition is not true, we complete the matrix $P$
by replacing the zero columns with $1/N$ (by the personalization vector
$V$ in a general case) to get $\overline{P}$.
This corresponds to the dangling nodes completion  \cite{deep},
we will see below that such an adaptation is not required in our
approach).

In this paper, we consider the iteration of equation of the form:
\begin{eqnarray}\label{eq:iteration}
 X_{n+1} = A . X_n
\end{eqnarray}
where $A$ is a matrix of size $N\times N$ which can be explicitly decomposed as:
\begin{eqnarray}\label{eq:doeb}
A &=& d P + (1-d) V.\ind^t
\end{eqnarray}
where $V$ is a normalized vector of size $N$
and $\ind^t$ is the column vector with all components equal to one.
The equation \eqref{eq:doeb} is an explicit Doeblin's decomposition \cite{doeb, doob}.
So we have:
\begin{eqnarray*}\label{eq:iteration2}
 X_{n+1} = d P X_n + (1-d) V.
\end{eqnarray*}

We assume that the stationary probability of $A$ is defined by the vector $X$
(its existence and uniqueness is straightforward e.g. from the contraction property of $A$).
We have:
$$
X = d P X + (1-d) V.
$$

In the context of PageRank, the vector $V$ is a personalized initial
condition cf. \cite{deep}.

\subsection{Main equations}
We first define the vector $S_n$ by:
$$
S_{n} = (1-d) \sum_{k=0}^{n} d^k P^k V
$$
So that:
$$
S_{\infty} = (1-d) \sum_{k=0}^{\infty} d^k P^k V = (1-d) (I_d-dP)^{-1} V = X
$$

The equation $S_n$ has been used in \cite{Boldi2009} to define formulae
for PageRank derivatives of any order.\\

We define $J_k$ a matrix with all entries equal to zero except for
the k-th diagonal term: $(J_k)_{kk} = 1$.

In the following, we assume given a deterministic or random sequence
$I = \{i_1, i_2, ..., i_n,...\}$ with $i_n \in \{1,..,N\}$.
We only require that the number of occurrence of each value $k\in \{1,..,N\}$
in $I$ to be infinity.

Then, we define the fluid vector $F$ associated to $I$ by:
\begin{eqnarray}
F_0 &=& (1-d) V\\
F_n &=& d P J_{i_n}F_{n-1} + \sum_{k\neq i_n}J_k F_{n-1}\\
&=& (I_d - J_{i_n} + dPJ_{i_n}) F_{n-1}\label{eq:defF}
\end{eqnarray}
where $I_d$ is the identity matrix.

We define the history vector $H$ by:
\begin{eqnarray}\label{eq:defH}
H_n &=& \sum_{k=1}^n J_{i_k} F_{k-1}.
\end{eqnarray}

By definition, $H_n$ is an increasing function (all components
are positive).

The above fluid and history vector is associated to the following
algorithm (ALGO-REF):

\begin{verbatim}
Initialization:
  H[i] := 0;
  F[i] := (1-d)*v_i;
  R := 1-d;

k := 1;
While ( R/(1-d) > Target_Error )
  Choose i_k;
  sent := F[i_k];
  F[i_k] := 0;
  H[i_k] += sent;
  For all child node j of i_k:
    F[j] += sent*p(j,i_k)*d;
  R -= sent*(1-d);
  k++;
\end{verbatim}

\begin{remark}
If $d=1$, the above algorithm is
equal to the one defined in \cite{serge}.
The role of $d$ is here important: in \cite{serge}, the stopping condition
is not defined, because in their algorithm, what they called cash (our fluid $F$) is not decreasing
but constant. Also, because in our algorithm, the total fluid tends to zero, 
we can predict and control precisely the convergence to the limit, as we will see
below.
\end{remark}

\section{Theoretical results}\label{sec:theory}
\subsection{Main equations}
\begin{theorem}\label{theo:main}
We have the equality:
\begin{eqnarray}\label{eq:H}
H_n + F_n &=& F_0 + d P H_n.
\end{eqnarray}
\end{theorem}

\proof
The proof is straightforward by induction: assuming the equation \eqref{eq:H} true for $n$:
\begin{eqnarray*}
H_{n+1} + F_{n+1} &=& H_n + J_{i_{n+1}} F_n + \sum_{k\neq i_{n+1}}J_k F_{n} + dP J_{i_{n+1}} F_n\\
&=& H_n + F_n + dP J_{i_{n+1}} F_n\\
&=& F_0 + dP H_n + dP J_{i_{n+1}} F_n\\
&=& F_0 + dP H_{n+1}
\end{eqnarray*}
Note that the equation \eqref{eq:H} is true for $d=1$ with $F_0$ replaced
by any initial vector.

\begin{theorem}\label{theorem:Sinf}
We have the equality:
\begin{eqnarray}\label{eq:S}
S_{\infty} - H_n &=& \sum_{k=0}^{\infty} d^k P^k F_n = ( I - d P )^{-1} F_n
\end{eqnarray}
and as a direct consequence, we have:
\begin{eqnarray}\label{eq:SL}
|S_{\infty} - H_n| &=& \frac{|F_n|}{1-d}
\end{eqnarray}
where $|\cdot|$ is the $L_1$ norm.
\end{theorem}

\proof
$$
S_{\infty} = d P S_{\infty} + F_0
$$
Using the equation \eqref{eq:H}, we have:
\begin{eqnarray*}
S_{\infty} - H_n &=& d P S_{\infty} + F_0 - ( F_0 + P H_n ) + F_n\\
&=& d P (S_{\infty} - H_n) + F_n.
\end{eqnarray*}
By iteration, we get \eqref{eq:S}. The equation \eqref{eq:SL} is obvious
remarking that $|P^k F_n| = |F_n|$

Now, we assume that the sequence $I$ is chosen such that at iteration $n$:
$i_n = \arg\max_i (F_{n-1})_i$. Then we have the following result:

\begin{lemma}\label{lemma}
If $i_n = \arg\max_i (F_{n-1})_i$, we have:
\begin{eqnarray*}
|F_{n}| &\le& |F_{n-1}| \left( 1 - \frac{1-d}{N} \right). 
\end{eqnarray*}
\end{lemma}
\proof
The proof is straightforward, noticing that we suppressed $J_{i_n}F_{n-1}$
and added $J_{i_n}F_{n-1} \times d$. Then using $J_{i_n}F_{n-1} \ge |F_{n-1}|/N$.

Thanks to Lemma \ref{lemma}, we have an exponential convergence to zero of $F_n$.
This lemma still holds if for $i_n$ an entry of $F_{n-1}$ which is larger or equal to $|F_{n-1}|/N$
is chosen.

As a consequence of this lemma, we have:
\begin{eqnarray*}
H_{\infty} &=& F_0 + d P H_{\infty}.
\end{eqnarray*}
Therefore, $H_{\infty} = X$ and $H_n$ is an increasing (component by component) function
to $X$.

\begin{remark}
From equations \eqref{eq:H} and \eqref{eq:defF}, we can obtain the following
iteration equation on $H$:
\begin{eqnarray*}
H_{n} &=& \left(I_d - J_{i_n}(I_d - dP)\right)H_{n-1} + J_{i_n} (1-d)V.
\end{eqnarray*}
This formulation may be directly exploited for an efficient distributed
computation of $H_{\infty}$. This issue will be addressed in a future paper.

If $d=1$, we still have:
\begin{eqnarray*}
H_n &=& \left(I_d - J_{i_n}(I_d - P)\right)H_{n-1} + J_{i_n} X_0\\
X_n &=& P X_{n-1}, 
\end{eqnarray*}
and $H_n$ converges to $\sum_{k=0}^{\infty}X_k = \sum_{k=0}^{\infty}P^k X_0$
after appropriate normalization (case of \cite{serge}), at least when all entries
of $P$ are non-negative (when the spectral radius of $P$ is one and if there are negative
entries in $P$, it may converge or not converge): 
but this convergence is based on a Cesaro
averaging which is known to be very slow (cf. results in Section \ref{sec:simu}).


In all cases, $H_n$ can be simply interpreted as a specific way to compute or to estimate
the power series $\sum_{k=0}^{\infty}P^k X_0$ (when there is convergence).
\end{remark}

\subsection{Updating equation}
The fact that $H_n$ converges to $S_{\infty}$ for any arbitrary choice
of the sequence $I$ can be also exploited to compute more efficiently the
new eigenvector in case of the graph modification (so of $P$).

\begin{theorem}\label{theo:update}
Assume the initial graph associated to $(P, H_{\infty})$ is modified to the updated graph 
represented by $(P',H_{\infty}')$, then we have:
\begin{eqnarray*}
H_{\infty}' - H_{\infty} &=& (1-dP')^{-1} d(P'-P) H_{\infty}.
\end{eqnarray*}
\end{theorem}
\proof
We have
\begin{eqnarray*}
(1-dP') (H_{\infty}' - H_{\infty}) &=& F_0 - (1-dP') H_{\infty}\\
&=& F_0 - (1-dP) H_{\infty} + d(P'-P) H_{\infty}\\
&=& d(P'-P) H_{\infty}.
\end{eqnarray*}

In the above updating equation, $(P'-P)H_{\infty}$ may mix positive and negative
terms. We can apply on them the operator $F_n'$ separately or jointly.

Now, assume the equation $H_n$ has been computed up to the iteration $n_0$ and that
at that time, we are interested to compute the limit associated to $P'$
(for instance, because the web graph has been modified/updated).

Then very naturally, we can apply our diffusion method with $P'$, but with modified
initial condition $F_0' = F_{n_0} + d(P' - P) H_{n_0}$ for which we have:
\begin{eqnarray}\label{eq:Hp}
H_n' + F_n' &=& F_0' + dP' H_n'.
\end{eqnarray}

We have the following very intuitive results:
\begin{theorem}
$H_{n_0} + H_{\infty}'$ ($H_{\infty}'$ is the limit of the equation \eqref{eq:Hp})
is the solution of the equation:
$$
X = d P' X + F_0.
$$
\end{theorem}
\proof
The limit of \eqref{eq:Hp} satisfies:
\begin{eqnarray*}
H_{\infty}' &=&  F_{0}' + dP' H_{\infty}'\\
            &=&  F_{n_0} + d(P' - P) H_{n_0} + dP' H_{\infty}'.
\end{eqnarray*}
Combining this with $H_{n_0} + F_{n_0} = F_0 + dP H_{n_0}$, we have obviously
what we want.

The above result implies that one can continuously update the iterations
when $P$ is regularly updated by just injecting in the {\em system} a fluid
quantity equal to $d(P' - P) H_{n_0}$ and then applying the new matrix $P'$.

If a distributed computation approach was to be used, we just need to synchronize
the time from which $P'$ is applied.

\subsection{About dangling nodes}
In the algorithm ALGO-REF, we don't need to complete the null columns of the matrix $P$.
We can simply add the following condition:

\begin{verbatim}
While ( R/(1-d) > Target_Error )
  Choose i_k;
  sent := F[i_k];
  H[i_k] += sent;
  F[i_k] := 0;
  If ( i_k has no child )
    R -= sent;
  else
    For all child node j of i_k:
      F[j] += sent*p(j,i_k)*d;
    R -= sent*(1-d);
  k++;
\end{verbatim}

The quantity $R/(1-d)$ measures exactly (thanks to Theorem \ref{theorem:Sinf})
the distance to the stationary probability. However when $P$ includes dangling
nodes, it is easy to see that $R/(1-d)$ defines only an upper bound and that
$H_{\infty}$ need to be renormalized (dividing by the norm $L_1$ of $H_{\infty}$) 
to find the probability eigenvector satisfying the PageRank equation with the
completed matrix $\overline{P}$.

\subsection{Asynchronous distributed computation}
The proposed algorithm is very well suited for the asynchronous distributed computation
(e.g. cf. \cite{jela}).
Indeed, at any moment of the iteration, 
the fluid $F_n$ can be split in any number of elements and be distributed per element,
the $L_1$ norm of each element controlling exactly the error that can be reduced.
The most natural way is to divide per component $(F_n)_i$ of the vector, and an obvious
particular example is to split the initial condition vector $X_0$ in
$N/m$ elements of size $m$.

\section{Comparison results}\label{sec:simu}
In the following, we will call:
\begin{itemize}
\item ALGO-MAX: if $I$ is defined by $i_n = \arg\max_i (F_{n-1})_i$;
\item ALGO-RAND: if $i_n$ is randomly chosen from $\{1,..,N\}$;
\item ALGO-PER: if $i_n = n \mod N$ (periodic cycle);
\item MAT-ITER: if the matrix product iteration \eqref{eq:iteration} is used;
\item OPIC: the one defined in \cite{serge} (Random version).
\end{itemize}
The stopping condition for ALGO-REF variants are based on $R/(1-d) >$ \verb+Target_Error+.
For MAT-ITER, we use the well known condition: $|X_{n}-X_{n-1}|\times d/(1-d) < $ \verb+Target_Error+.
For OPIC, we measured the convergence by comparing the distance to the precomputed limit
(from MAT-ITER), since it does not define any stopping condition.

Below we set a simple simulation scenario to get a first evaluation of our
proposed solution and comparison to existing solutions in the
context of the original PageRank on the web graph.
We don't pretend to generate any realistic model, for more details on the
web graph the readers may refer to \cite{broder, kumar, barabasi, barabasi00, levene}.

\subsection{Simulation scenario}
We set $N$ the total number of nodes (URLs) to be simulated.
Then we create $L$ random links (directional) to connect a node $i$
to $j$ as follow:

\begin{itemize}
\item the choice of the source node is done following a power-law: $1/k^\alpha$;
\item the choice of the destination node is done following a power-law: $1/k^\alpha$.
\end{itemize}

For simplicity, we assumed no correlation between the number of incoming and outgoing
links: for that purpose,
to defined the distribution of the destination nodes by
associating to node $k$ a probability proportional to $1/k^\alpha$
followed by a large number (by default $N$) of permutations of randomly chosen pair 
of nodes $(i,j)$.
Then, to define the distribution of the destination nodes,
we did the same operation. In this way, the order of the nodes does not introduce
any bias for ALGO-PER.

The tables below summarize the characteristics of the 6 synthetic data
we considered varying the number of links per node and the parameter $\alpha$.

\begin{table}
\begin{center}
\begin{tabular}{|l|cc|}
\hline
Scenario & L (nb links) & D (Nb dangling nodes)\\
\hline
S1 & 2172 & 9552\\
S2 & 8081 & 8646\\
S3 & 28507 & 6252\\
\hline
\end{tabular}\caption{Parameters: $\alpha=2.0$, $N=10000$}
\end{center}
\end{table}

\begin{table}
\begin{center}
\begin{tabular}{|l|cc|}
\hline
Scenario & L (nb links) & Nb dangling nodes\\
\hline
S1b & 12624 & 7696\\
S2b & 61189 & 3197\\
S3b & 265245 & 33\\
\hline
\end{tabular}\caption{Parameters: $\alpha=1.5$, $N=10000$}
\end{center}
\end{table}

\subsection{Simulation results analysis}
We define the elementary step as an operation requiring the use of one non-zero
entry of $A$: for instance, if $A$ has $L$ entries that are not zero,
the product $A.X$ would require $L$ elementary steps.
We already mentioned that ALGO-REF variants does not require the matrix completion $\overline{P}$.
For MAT-ITER, we can also avoid such a completion, but the cost to pay is the vector
renormalization. Therefore, we assumed here that the cost of $A.X$ is $L$ where $L$ is the
number of not zero entries of the initial matrix $P$.


For ALGO-RAND and ALGO-PER, when we have no fluid on the node $i_n$ ($F_{i_n} = 0$),
we go to the next step without incrementing the counter of the
elementary steps.

For ALGO-MAX, there is of course a computation cost to find the $\arg\max$ of $F_n$.
In order to show its full potential, its cost is not taken into account in the 
number of elementary steps. In the comparison tests, we found that taking the
argmax is not necessary and same improvements were obtained replacing the argmax
by iteration process where in one iteration all $i$ such that $(F_n)_i$ is above
a threshold are all chosen, then scaling down the threshold progressively.

Figure \ref{fig:S1}, \ref{fig:S2} and \ref{fig:S3} shows the results for S1, S2 and S3.
In these scenarios, because of $\alpha$, there are the proportion of the dangling
nodes are important. We can observe in all cases a substantial gain with our
approach. 

\begin{figure}[htbp]
\centering
\includegraphics[angle=-90,width=\linewidth]{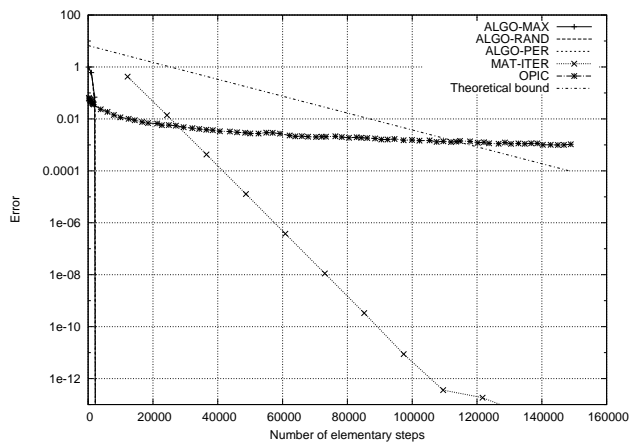}
\caption{Scenario S1. $L/N = 0.22$, $D/N = 0.96$.}
\label{fig:S1}
\end{figure}

\begin{figure}[htbp]
\centering
\includegraphics[angle=-90,width=\linewidth]{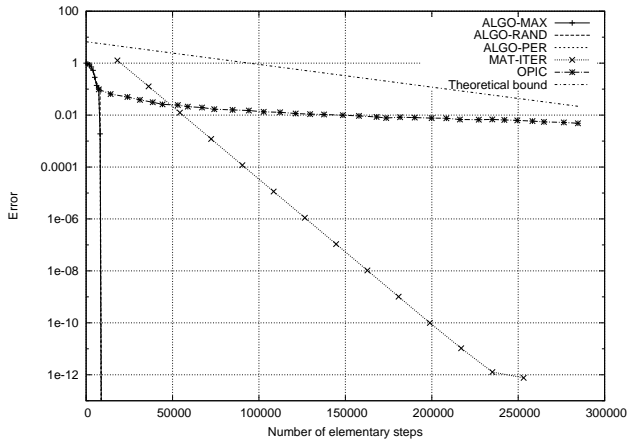}
\caption{Scenario S2. $L/N = 0.81$, $D/N = 0.86$.}
\label{fig:S2}
\end{figure}

\begin{figure}[htbp]
\centering
\includegraphics[angle=-90,width=\linewidth]{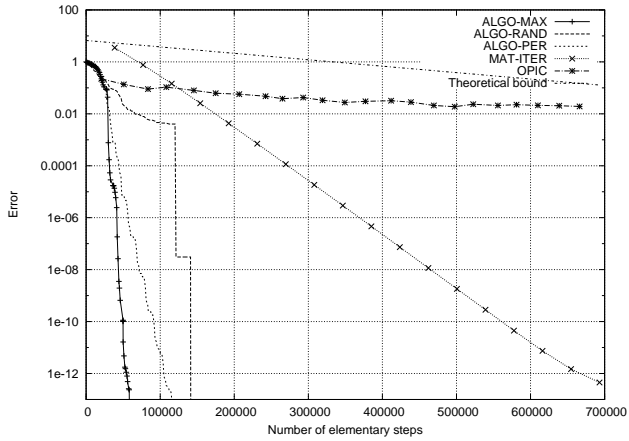}
\caption{Scenario S3. $L/N = 2.9$, $D/N = 0.62$.}
\label{fig:S3}
\end{figure}

Figure \ref{fig:S1b}, \ref{fig:S2b} and \ref{fig:S3b} shows the results for S1b, S2b and S3b.
In these scenarios, the proportion of the dangling nodes are less important. 
We can still observe in all cases a substantial gain with our
approach. 

\begin{figure}[htbp]
\centering
\includegraphics[angle=-90,width=\linewidth]{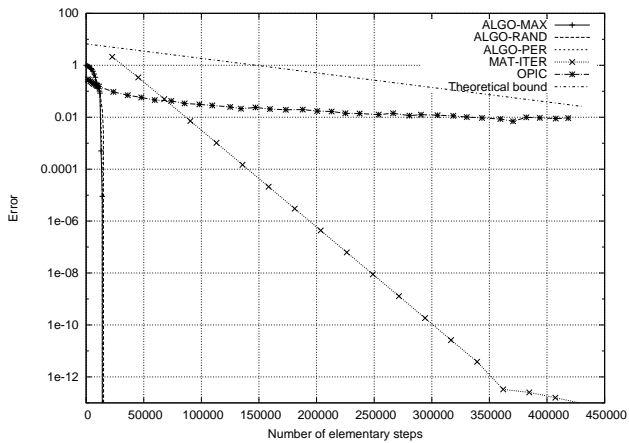}
\caption{Scenario S1b. $L/N = 1.3$, $D/N = 0.77$.}
\label{fig:S1b}
\end{figure}

\begin{figure}[htbp]
\centering
\includegraphics[angle=-90,width=\linewidth]{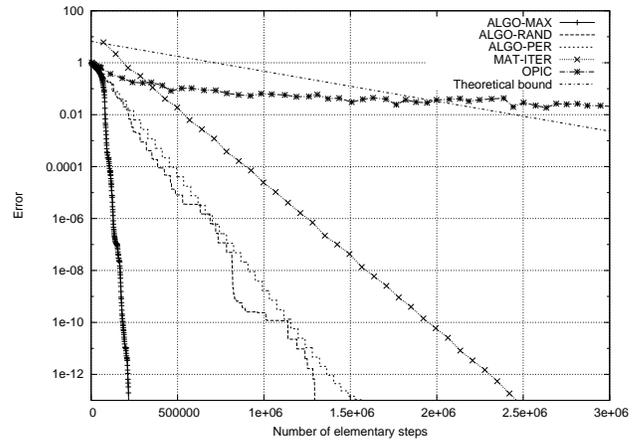}
\caption{Scenario S2b. $L/N = 6.1$, $D/N = 0.32$.}
\label{fig:S2b}
\end{figure}

\begin{figure}[htbp]
\centering
\includegraphics[angle=-90,width=\linewidth]{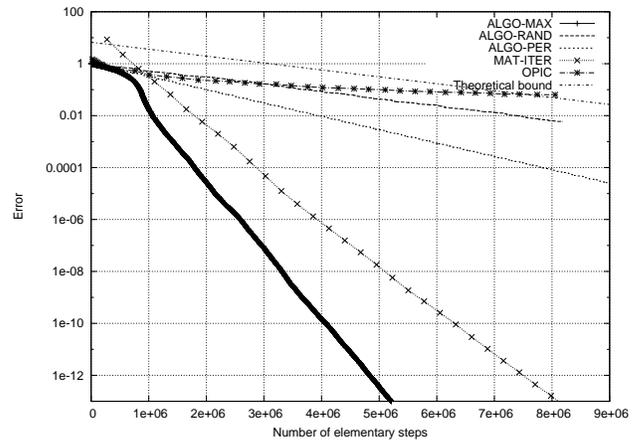}
\caption{Scenario S3b. $L/N = 26.5$, $D/N = 0.003$.}
\label{fig:S3b}
\end{figure}

\subsection{Comparison on web graph}
In this section, we realized the same comparison tests than above on a
web graph imported from the dataset \verb+gr0.California+
(available on \verb+http://www.cs.cornell.edu/Courses/cs685/+ \verb+2002fa/+)
and from the dataset \verb+uk-2007-05@1000000+
(available on \verb+http://law.dsi.unimi.it/datasets.php+).

The results for the dataset \verb+gr0.California+ is shown in Figure
\ref{fig:dataset1}. In this figure, we added ALGO-OP a variant of ALGO-MAX
where the argmax is replaced by 
$$i_n = \arg\max_i \left((F_{n-1})_i/((\#in_i +1) \times (\#out_i+1))\right)$$
with $\#in_i$ (resp. $\#out_i$) equal to the number of incoming (resp. outgoing) 
links to (resp. from) the node $i$. The intuition of this renormalization is:
\begin{itemize}
\item \#out: the cost of the diffusion of the fluid $F$ is proportional to \#out;
\item \#in: when there are many incoming links, it is worth to wait and aggregate
  the fluid before the fluid diffusion.
\end{itemize}
We see that we still have a significant gain and that ALGO-OP outperforms ALGO-MAX.
The main reason for which the proposed approach outperforms greatly the original
OPIC algorithm is the fact that we don't have the fluctuation due to
the Cesaro averaging as in OPIC (which converges as $1/sqrt(n)$).

\begin{figure}[htbp]
\centering
\includegraphics[angle=-90,width=\linewidth]{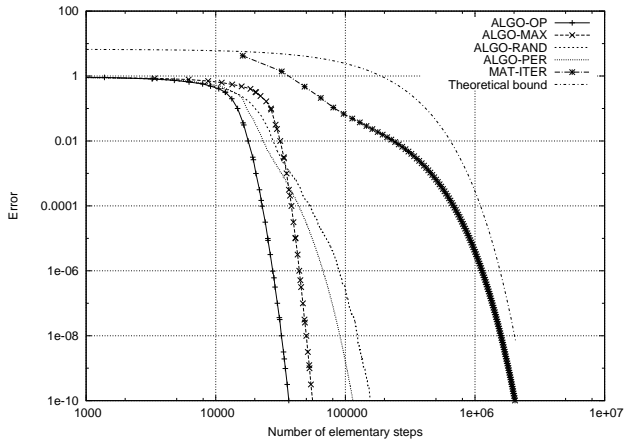}
\caption{Dataset gr0.California: 16150 links on 9664 nodes (4637 dangling nodes).}
\label{fig:dataset1}
\end{figure}

Figure \ref{fig:dataset2} shows the results for the dataset \verb+uk-2007-05@1000000+.
Here ALGO-MAX outperforms ALGO-OP. It shows that the optimization of the sequence choice $I$ is still
an open problem. We added here ALGO-OP2 based on:
$$i_n = \arg\max_i \left((F_{n-1})_i/(\#out_i+1)\right).$$

\begin{figure}[htbp]
\centering
\includegraphics[angle=-90,width=\linewidth]{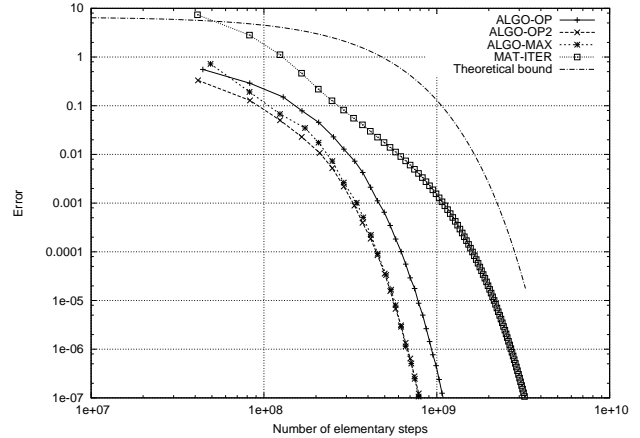}
\caption{Dataset uk-2007-05@1000000: 41247159 links on 1000000 nodes (45766 dangling nodes).}
\label{fig:dataset2}
\end{figure}

\section{Conclusion}\label{sec:conclusion}
In this paper, we proposed a new algorithm to optimize the
computation cost of the eigenvector of PageRank equation and showed
the theoretical potential gain.

We believe that we have here a promising new approach and we are
still investigating on a practical optimal ALGO-OP variants solution.

In a future work, we wish to validate the performance of our approach
in terms of the effective run time cost for large data, in particular,
using the asynchronous distributed computation approach.

\section*{Acknowledgments}
The author is very grateful to Fran\c cois Baccelli for his very valuable comments
and suggestions.

\end{psfrags}
\bibliographystyle{abbrv}
\bibliography{sigproc}

\end{document}